\documentclass[aps,prb,twocolumn,groupedaddress,amsmath,showpacs]{revtex4}
    \usepackage{bm}
    \usepackage{graphicx}
\newcommand\beq{\begin{equation}}
\newcommand\eeq{\end{equation}}
\newcommand\beqa{\begin{eqnarray}}
\newcommand\eeqa{\end{eqnarray}}
\newcommand{\dd}{\text{d}}
\newcommand{\nn}{\nonumber\\}
\newcommand{\hs}{\text{HS}}
\newcommand{\eff}{\text{eff}}
\newcommand{\Ss}{\text{SS}}

\begin{document}
\title{On the equivalence between the energy and virial routes to the equation of state of hard-sphere fluids}
\author{Andr\'es Santos}
\email{andres@unex.es}
\homepage{http://www.unex.es/eweb/fisteor/andres/}
\affiliation{Departamento
de F\'{\i}sica, Universidad de Extremadura, E-06071 Badajoz, Spain}
\date{\today}
\begin{abstract}
The energy route to the equation of state of hard-sphere fluids is
ill-defined since the internal energy is just that of an ideal gas
and thus it is independent of  density.  It is shown that this
ambiguity can be  avoided by considering a square-shoulder
interaction and taking the limit of vanishing shoulder width. The
resulting hard-sphere equation of state coincides exactly with the
one obtained through the virial route. {}Therefore, the energy and
virial routes to the equation of state of hard-sphere fluids can be
considered as equivalent.

\end{abstract}
\maketitle

\maketitle

Given a fluid of particles interacting via a two-body potential
$\varphi(r)$, its equation of state (EOS) can be obtained in terms
of the radial distribution function $g(r;\rho,\beta)$, where $\rho$
is the number density and $\beta=1/k_BT$ is the inverse temperature,
through a number of routes.\cite{HM86} The most common ones are the
virial route,
\beq
\frac{\beta p}{\rho}\equiv
Z(\rho,\beta)=1+2^{d-1}v_d\rho\int_0^\infty \dd r\, r^d
y(r;\rho,\beta)\frac{\partial}{\partial r} e^{-\beta \varphi(r)},
\label{1}
\eeq
the compressibility route
\beqa
\left(\beta\frac{\partial p}{\partial \rho}\right)^{-1}\equiv
\chi(\rho,\beta)&=&1+2^{d}dv_d\rho\int_0^\infty \dd r\, r^{d-1}\nn
&& \times\left[h(r;\rho,\beta)-1\right],
\label{2}
\eeqa
and the energy route
\beqa
u(\rho,\beta)&=&\frac{d}{2\beta}\left[1+2^{d}v_d\rho\beta\int_0^\infty
\dd r\, r^{d-1}\varphi(r)e^{-\beta \varphi(r)}\right.\nn
&&\left.\times y(r;\rho,\beta)\right].
\label{3}
\eeqa
In Eqs.\ (\ref{1})--(\ref{3}), $p$ is the pressure, $Z$ is the
compressibility factor, $d$ is the dimensionality of the system,
$v_d=(\pi/4)^{d/2}/\Gamma(1+d/2)$ is the volume of a $d$-dimensional
sphere of unit diameter, $\chi$ is the isothermal susceptibility,
$u$ is the internal energy per particle, $y(r)\equiv
\exp[\beta\varphi(r)]g(r)$ is the cavity function, and $h(r)\equiv
g(r)-1$ is the total correlation function. If the \textit{exact}
function $g(r;\rho,\beta)$ is inserted, the three routes are
thermodynamically consistent, i.e.,
\beq
\chi^{-1}(\rho,\beta)=\frac{\partial}{\partial\rho}\left[\rho
Z(\rho,\beta)\right],
\label{4}
\eeq
\beq
\rho\frac{\partial}{\partial\rho}u(\rho,\beta)=\frac{\partial}{\partial
\beta}{Z(\rho,\beta)}.
\label{5}
\eeq
On the other hand, if an \textit{approximate} function
$g(r;\rho,\beta)$ is used, the compressibility factor obtained
directly from Eq.\ (\ref{1}) does not necessarily coincide with that
obtained from the combination of Eqs.\ (\ref{2}) and (\ref{4}) or
with that obtained from the combination of Eqs.\ (\ref{3}) and
(\ref{5}).

Let us now particularize to the hard-sphere (HS) interaction
\beq
\varphi_\hs(r)=\begin{cases} \infty,&r<\sigma,\\
0,&r>\sigma.
\end{cases}
\label{6}
\eeq
In that case, $y(r;\rho,\beta)=y_\hs(r;\rho\sigma^d)$ and Eq.\
(\ref{1}) becomes
\beq
Z_\hs(\rho\sigma^d)=1+2^{d-1}v_d\rho\sigma^d
y_\hs(\sigma;\rho\sigma^d).
\label{7}
\eeq
On the other hand, Eq.\ (\ref{3}) reduces to
\beq
u_\hs(\beta)=\frac{d}{2\beta}.
\label{8}
\eeq
Since $Z_\hs$ is independent of temperature and $u_\hs$ is
independent of density, Eq.\ (\ref{5}) is trivially satisfied as
$0=0$ and so it is not possible in principle to get the
compressibility factor from the internal energy.

A way of circumventing the ill-definition of the energy route to the
EOS of an HS fluid consists of considering a convenient interaction
potential that encompasses  that of hard spheres in certain limits
but for which the temperature plays a relevant role. The simplest
choice for such a potential is perhaps the so-called square-shoulder
(SS) potential:\cite{SS}
\beq
\varphi_\Ss(r)=\begin{cases} \infty,&r<\sigma,\\
\epsilon,&\sigma<r<\sigma',\\
0,&r>\sigma',
\end{cases}
\label{9}
\eeq
where $\epsilon$ is a positive constant. For this potential, Eq.\
(\ref{3}) becomes
\beq
u_\Ss(\rho,\beta)=\frac{d}{2\beta}\left[1+2^{d}v_d\rho\beta\epsilon
e^{-\beta\epsilon}\int_\sigma^{\sigma'} \dd r\, r^{d-1}
y_\Ss(r;\rho,\beta)\right].
\label{10}
\eeq
The SS interaction is equivalent to an HS interaction of diameter
$\sigma$ in the infinite-temperature limit ($\beta\epsilon\to 0$)
and to an HS interaction of diameter $\sigma'$ in the
zero-temperature limit ($\beta\epsilon\to \infty$). Of course, the
equivalence $\text{SS}\to\text{HS}$ also holds in the limit of zero
shoulder width ($\sigma'\to\sigma$).

Let us  now imagine that  $y_\Ss(r;\rho,\beta)$ is known, either
exactly or approximately (e.g., as obtained from  the Percus--Yevick
approximation), and thus it is possible to compute
$u_\Ss(\rho,\beta)$ from Eq.\ (\ref{10}). Then, we can get
$Z_\Ss(\rho,\beta)$ from Eq.\ (\ref{5}) as
\beqa
Z_\Ss(\rho,\beta)-Z_\hs(\rho\sigma^d)&=&2^{d-1}dv_d\epsilon\rho\frac{\partial}{\partial\rho}\rho
\int_0^\beta\dd\beta'\, e^{-\beta'\epsilon}\nn &&\times
\int_\sigma^{\sigma'} \dd r\, r^{d-1}y_\Ss(r;\rho,\beta'),\nn
\label{11}
\eeqa
where we have made use of the property $\lim_{\beta\epsilon\to
0}Z_\Ss(\rho,\beta)=Z_\hs(\rho\sigma^d)$. Taking into account the
complementary condition $\lim_{\beta\epsilon\to
\infty}Z_\Ss(\rho,\beta)=Z_\hs(\rho{\sigma'}^d)$, Eq.\ (\ref{11})
yields
\beqa
\frac{Z_\hs(\rho{\sigma'}^d)-Z_\hs(\rho\sigma^d)}{\rho{\sigma'}^d-\rho{\sigma}^d}&=&
\frac{2^{d-1}dv_d\epsilon}{{\sigma'}^d-{\sigma}^d}\frac{\partial}{\partial\rho}\rho
\int_0^\infty\dd\beta\, e^{-\beta\epsilon}\nn &&\times
\int_\sigma^{\sigma'} \dd r\, r^{d-1}y_\Ss(r;\rho,\beta). \nn &&
\label{12}
\eeqa

Equation (\ref{12}) can be considered as the condition defining the
compressibility factor of an HS fluid associated with the energy
route. To proceed further, we take the limit $\sigma'\to\sigma$.
Thus,
\beq
\lim_{\sigma'\to\sigma}\frac{Z_\hs(\rho{\sigma'}^d)-Z_\hs(\rho\sigma^d)}{\rho{\sigma'}^d-\rho{\sigma}^d}
=\sigma^{-d}\frac{\partial}{\partial\rho}Z_\hs(\rho\sigma^d),
\label{13}
\eeq
\beq
\lim_{\sigma'\to\sigma}\frac{1}{{\sigma'}^d-{\sigma}^d}\int_\sigma^{\sigma'}
\dd r\,
r^{d-1}y_\Ss(r;\rho,\beta)=\frac{1}{d}y_\hs(\sigma;\rho\sigma^d).
\label{13bis}
\eeq
Therefore, Eq.\ (\ref{12}) reduces in the limit $\sigma'\to\sigma$
to
\beq
\frac{\partial}{\partial\rho}Z_\hs(\rho\sigma^d)=2^{d-1}v_d
\frac{\partial}{\partial\rho}\rho\sigma^d
y_\hs(\sigma;\rho\sigma^d).
\label{14}
\eeq
Finally, integrating over density and imposing the ideal gas
boundary condition $Z_\hs(0)=1$, the virial EOS (\ref{7}) is
reobtained.

 The
generalization to mixtures (either additive or non-additive) is
straightforward. In that case, the SS potential for each pair $ij$
is defined by Eq.\ (\ref{9}) with the changes
$\epsilon\to\epsilon_{ij}$, $\sigma\to\sigma_{ij}$, and
$\sigma'\to\sigma_{ij}'=\lambda\sigma_{ij}$, where the factor
$\lambda$ is common to all the pairs. The equation equivalent to
Eq.\ (\ref{12}) in the case of mixtures is
\beqa
\frac{Z_\hs(\rho{\sigma'}_\eff^d)-Z_\hs(\rho\sigma_\eff^d)}{\rho(\lambda^d-1)}=
\frac{2^{d-1}dv_d}{\lambda^d-1}\frac{\partial}{\partial\rho}\rho\sum_{ij}x_ix_j\nn
\times \epsilon_{ij} \int_0^\infty\dd\beta\,
e^{-\beta\epsilon_{ij}}\int_{\sigma_{ij}}^{\sigma_{ij}'} \dd r\,
r^{d-1}y^\Ss_{ij}(r;\rho,\beta),
\label{15}
\eeqa
where $\{x_i\}$ are mole fractions and
$\sigma_\eff^d\equiv\sum_{i,j}x_ix_j \sigma_{ij}^d$,
${\sigma'}_\eff^d\equiv\sum_{i,j}x_ix_j {\sigma'}_{ij}^d=\lambda^d
\sigma_\eff^d$. Taking the limit $\lambda\to 1$ in Eq.\ (\ref{15})
one gets
\beq
Z_\hs(\rho\sigma_\eff^d)=1+2^{d-1}v_d\rho\sum_{ij}x_ix_j\sigma_{ij}^d
y^\hs_{ij}(\sigma_{ij};\rho\sigma_\eff^d),
\label{16}
\eeq
which is not but the virial EOS for an HS mixture.

In summary, in this note I have shown that the ill-definition of the
energy route to the EOS of an HS fluid is saved by first considering
an SS fluid and then taking the limit of vanishing shoulder width.
The resulting EOS coincides exactly with the one obtained through
the virial route. {}From that point of view, the energy and virial
routes to the EOS of HS fluids can be considered as equivalent.

\acknowledgments
 This research
has been supported by the Ministerio de Educaci\'on y Ciencia
(Spain) through Grant No. FIS2004-01399 (partially financed by FEDER
funds).

\end{document}